\documentclass[pre,showpacs,superscriptaddress,twocolumn,longbibliography]{revtex4-1}

\usepackage{hyperref}
\usepackage{color}
\usepackage[usenames,dvipsnames]{xcolor}
\usepackage{amsmath,amsthm,amssymb}
\usepackage{graphicx}
\usepackage{float}
\usepackage{epsfig}
\usepackage{bm}% bold math
\usepackage{mathrsfs}
\usepackage{multirow}
\usepackage[all]{xy}
\usepackage{pbox}
\usepackage{verbatim}
\usepackage{braket}
\usepackage{mathtools}
\usepackage{bm}
\usepackage{tikz}
\usepackage{xcolor}
 
\usepackage{mathtools}

\usepackage{mathtools}
\usepackage{tabstackengine}
\usepackage{enumerate}   
\usepackage[normalem]{ulem}
\stackMath
\DeclareMathOperator{\Tr}{Tr}
\DeclareMathOperator{\Var}{Var}
\newcommand{\er}[1]{Eq.~\eqref{#1}}
\newcommand{\ers}[2]{Eqs.~(\ref{#1}-\ref{#2})}
\newcommand{\era}[2]{Eqs.~(\ref{#1}) and (\ref{#2})}

\newcommand{\beq}{\begin{equation}}
\newcommand{\eeq}{\end{equation}}

\newcommand{\W}{\mathbb W}
\renewcommand{\P}{\mathbb P}

\begin{document}  

\title{Optimal sampling of dynamical large deviations via matrix product states}

\author{Luke Causer}
\affiliation{School of Physics and Astronomy, University of Nottingham, Nottingham, NG7 2RD, UK}
\affiliation{Centre for the Mathematics and Theoretical Physics of Quantum Non-Equilibrium Systems,
University of Nottingham, Nottingham, NG7 2RD, UK}
\author{Mari Carmen Ba\~nuls}
\affiliation{Max-Planck-Institut f\"ur Quantenoptik, Hans-Kopfermann-Str.\ 1, D-85748 Garching, Germany}
\affiliation{Munich Center for Quantum Science and Technology (MCQST), Schellingstr.\ 4, D-80799 M\"unchen}
\author{Juan P. Garrahan}
\affiliation{School of Physics and Astronomy, University of Nottingham, Nottingham, NG7 2RD, UK}
\affiliation{Centre for the Mathematics and Theoretical Physics of Quantum Non-Equilibrium Systems,
University of Nottingham, Nottingham, NG7 2RD, UK}
\affiliation{All Souls College, Oxford, UK}

\begin{abstract}
The large deviation (LD) statistics of dynamical observables is encoded in the spectral properties of deformed Markov generators. Recent works have shown that tensor network methods are well suited to compute the relevant leading eigenvalues and eigenvectors accurately. However, the efficient generation of the corresponding rare trajectories is a harder task. Here we show how to exploit the MPS approximation of the dominant eigenvector to implement an efficient sampling scheme which closely resembles the optimal (so-called ``Doob'') dynamics that realises the rare events. We demonstrate our approach on three well-studied lattice models, the Fredrickson-Andersen and East kinetically constrained models (KCMs), and the symmetric simple exclusion process (SSEP). We discuss how to generalise our approach to higher dimensions.
\end{abstract}

\maketitle

\section{Introduction}

The complex behaviour of the non-equilibrium dynamics of stochastic systems can be characterised by studying trajectory ensembles, that is, the set of all possible trajectories alongside the probability that they occur under the evolution defined via a stochastic master operator.
This is analogous to standard thermodynamics, where static properties are entirely determined by the equilibrium ensemble of all microstates and the probabilities~\cite{Chandler1987}.
Often dynamical behaviour of interest is dominated not by trajectories that are typical under the dynamics, but by ``rare events'' which are exponentially (in time and in system size) scarce. 
Studying these rare events is made possible by using the framework of large deviations (LDs) \cite{Touchette2009, Garrahan2007, Lecomte2007, Garrahan2009, Garrahan2018, Jack2020}, where in large time limits time-extensive dynamical observables obey a LD principle, and their statistics is encoded in functions which play for dynamics the role that thermodynamic potentials play for statics  
(see below for definitions).

LD functions can be obtained in principle from a deformation or {\em tilting} of the dynamical generator (in the case of continuous-time dynamics) or the Markov matrix (in the case of discrete-time dynamics), through its largest eigenvalue. Obtaining this eigenvalue is not always an easy - or even possible - task, and often one needs to resort to numerical methods. Methods to overcome this difficulty often include techniques based on population dynamics, namely cloning or splitting \cite{Giardina2006, Lecomte2007b, Cerou2019, perez-espigares2019}, and importance sampling \cite{Bolhuis2002, Ray2018, Ray2018b, Klymko2018, Guyader2020} which provide information about the configurations frequently visited by the rare events. Notice that even if one manages to diagonalise the tilted generator (or the Markov matrix), the generation of rare trajectories is non-trivial: while rare trajectories are ``generated'' by the tilted operator, this is not a proper stochastic operator and these trajectories cannot be directly sampled. 

The efficient sampling of rare events can be achieved by searching for another stochastic dynamics which generates trajectories with desirable probabilities that are the same as (or a close approximation to) those of the tilted generator (with any small discrepancy corrected via importance sampling techniques). Methods for doing so currently include optimal control \cite{Nemoto2016, Ferre2018} and machine learning approaches, where one attempts to ``learn'' this convenient sampling dynamics \cite{oakes2020, rose2020, gillman2020}.
The optimal choice for a reference dynamics is the so-called {\em generalised Doob dynamics} \cite{Simon2009, Popkov2010, Jack2010, Chetrite2015, Garrahan2016, Carollo2018}, which generates trajectories with the exact tilting corresponding to the deformed generator. The Doob dynamics thus produces rare trajectories of the original dynamics ``on demand''. To construct such optimal dynamics, however, requires knowledge of the leading eigenvector of the tilted generator.

Variational tensor network (TN) techniques \cite{Vidal2003, Verstraete2004, Schollwoeck2011, orus2019, Schuch2008, McCulloch2007, pirvu2010}, originally devised as a tool to study quantum many-body systems,
are also convenient for studying classical statistical systems \cite{Honecker1997, Yasuhiro1998, Kemper2002, Ueda2005}.
More recently, they have been shown to be useful in the context of LDs in stochastic dynamics \cite{Gorissen2009, Banuls2019, Helms2019, Helms2020, Causer2020}.
In particular, it is often both possible and easy to approximate the leading eigenstate of the tilted generator of a one-dimensional stochastic lattice system using a matrix product state (MPS) ansatz, even those with dynamical (i.e.\ LD) phase transitions. Recent works have made use of this eigenstate to determine the statistical properties of the dynamics \cite{Gorissen2009, Banuls2019, Helms2019, Helms2020, Causer2020}. To our knowledge, however, such TN approach has not been exploited yet to sample efficiently rare trajectories. This is what we do in this paper. We present a scheme to use the MPS approximation to the leading eigenvalue of the tilted generator to construct a new dynamics which very closely resembles the optimal Doob dynamics, and we show how we can use this new dynamics to efficiently sample rare events.

We focus on three paradigmatic models. The first two correspond to kinetically constrained models (KCMs) \cite{Ritort2003,Garrahan2011,Garrahan2018}, specifically  
the Fredrickson-Andersen (FA) and the East \cite{Jackle1991} model, two well-studied models known for their connection to structural glasses \cite{Chandler2010, Biroli2013}. The third model is the symmetric simple exclusion processes (SSEP) \cite{Mallick2015, Blythe2007}. All these models have interesting LD statistics, including trajectory phase transitions controlled by their activities and/or currents (in the case of the SSEP) \cite{Bodineau2007,Garrahan2007,Appert-Rolland2008,Garrahan2009,Bodineau2012,Jack2015}.

The paper is organised as follows. In Sec.~II, we review continuous time Markov dynamics and LDs. We also recap how one can apply an MPS ansatz to study KCMs.
In Sec.~III, we define the Doob dynamics and introduce a scheme to approximate it with a reference dynamics, constructed using an MPS approximation to the leading eigenstate of the tilted generator.
In Sec.~IV we present the numerical results from our method applied to the three models. We show how our approach can effectively be used to accurately measure the statistics of time-extensive observables. We provide an outlook on possible generalisations and our conclusions in Sec.~V.

\section{Large deviations and matrix product states}
\label{sec_LD}
In this section we introduce continuous-time Markov dynamics, giving specific examples in the context of kinetically constrained models (KCMs) and exclusion processes. We then also review the framework of large deviations (LDs) and how variational matrix product states (MPS) can be used to determine the LD statistics.

\subsection{Continuous time Markov dynamics for KCMs and exclusion processes}
We consider stochastic Markov dynamics which evolves continuously in time. Suppose we have some system with the set of configurations $\{x_{1}, x_{2}, \dots, x_{M}\}$ where $M$ is the size of the configuration space. The probability that the system is in some configuration $x$ at the time $t$ is encoded in the probability vector $\ket{P(t)} = \sum_{x} P(x, t)\ket{x}$ which evolves under the stochastic master equation
\beq
    \frac{d}{dt}\ket{P(t)} = \W \ket{P(t)} \text{.}
\eeq
Here the generator of the dynamics $\W$ is given by
\beq
    \W = \sum_{x, x'\neq x} w_{x\to x'}\ket{x'}\bra{x} - \sum_{x} R_{x}\ket{x}\bra{x} \text{,}
    \label{W}
\eeq
where $w_{x\to x'}$ are the transition rates from configuration $x$ to $x'$ and $R_{x} = \sum_{x'\neq x} w_{x\to x'}$ is the escape rate from $x$. The largest eigenvalue of the generator is zero, with left eigenvector the flat state $\bra{-} = \sum_{x}\bra{x}$, and right eigenvector
%the flat state $\bra{-} = \sum_{x}\bra{x}$ as the left eigenvector and 
the steady state $\ket{\rm ss} = \sum_{x}P(x)\ket{x}$, which describes the probability of finding any configuration at equilibrium. If our system obeys detailed balance, then we are guaranteed that any initial state will eventually relax to some equilibrium state given enough time. Here we assume this to be the case.

We will focus on two broad areas of 1D constrained systems. The first is KCMs (for reviews see \cite{Ritort2003, Garrahan2011, Garrahan2018}), for which configuration changes are governed by a kinetic constraint which is explicitly encoded in the generator.
For concreteness, we focus on the 1D spin facilitation Fredrickson-Andersen (FA) \cite{Fredrickson1984} and East \cite{Jackle1991} models.
Both models are defined on a 1D lattice of $N$ binary variables (spins) $n_{j} = 0, 1$ for $j = 1, \dots, N$, and configuration changes are only allowed via single-spin flips.  The Markovian generators for both models are given by
\begin{align}
\W^{\text{East/FA}} = \sum_{i=1}^{N} \P_{i}^{\text{East/FA}} \big[ & c\sigma_{i}^{+} + (1 - c) \sigma_{i}^{-}
\label{Wkmc}
\\
& - c(1-n_{i}) - (1-c)n_{i}\big] \nonumber
\end{align}
where $\sigma_{i}^{\pm}$ are the Pauli raising/lowering operators acting on site $i$ and $c\in(0, 0.5]$ controls the rates at which spins flip, given they satisfy the kinetic constraints
\beq
    \quad \P_{i}^{\text{FA}} = n_{i-1} + n_{i+1} \text{,} \quad \P_{i}^{\text{East}} = n_{i-1},
    \label{constraints}
\eeq
where the first only allows a transition if the spin attempting to flip has a neighbouring excitation, and the second only if the neighbouring spin to the left is excited. (For the FA model the constraint is sometimes defined as the projector $n_{i-1} + n_{i+1} - n_{i-1} n_{i+1}$, but in practice it makes little difference with the definition above.)

The second area we consider are exclusion processes \cite{Blythe2007, Mallick2015} - particles hopping around sites on a lattice, with a hardcore exclusion such that we can have at most one particle per site. 
We focus on the 1D symmetric simple exclusion process (SSEP), adopting the lattice notation we used for KCMs, where now $n_{j} = 1 (0)$ implies the site is occupied (empty).
In the SSEP, a particle can hop left or right to its neighbouring sites, both with the same rate ($\gamma = 1/2$) if the neighbouring site is not already occupied.
The generator for the dynamics is
\beq
    \W^{\text{SSEP}} = \frac{1}{4}\sum_{i=1}^{N} \left(\sigma_{i}^{x}\sigma_{i+1}^{x} + \sigma_{i}^{y}\sigma_{i+1}^{y}+ \sigma_{i}^{z}\sigma_{i+1}^{z} - 1 \right)
    \label{WSSEP}
\eeq
where $\sigma^{a}_{i}$ are the Pauli operators acting on site $i$.

For the entirety of this paper, we will assume open boundary conditions (OBC), which will later reduce the computational cost of tensor network contractions. This formally means that we set $n_{0}=n_{N+1}=0$.
Furthermore, we impose certain restrictions on the state space. For the FA model, we simply exclude the disconnected zero state $n_{i} = 0, \ \forall i$.
On the other-hand, we set $n_{1} = 1$ for the East model which ensures the state space remains fully connected on each dynamical site $i > 1$.
Finally, we restrict SSEP such that the total number of particles $N_{p} = \sum_{i}n_{i}$ is fixed, with particle density $n_{p} = N_{p} / N$ which will be assumed to be $n_{p} = 1/2$.

\subsection{Trajectories and large deviations}

Consider some general trajectory $\omega_{t} = \{x_{0} \to x_{t_{1}} \to \dots \to x_{t_{K}}\}$ where the system moves into the configuration $x_{t_{i}}$ at time $t_{i}$ and has the total time $t > t_{K}$.
The dynamical activity $\hat{K}$ \cite{Lecomte2007, Garrahan2007, Garrahan2009, Garrahan2018, Maes2020} is a trajectory observable which measures the number of configuration changes for a given trajectory.
The probability of observing some activity $K$ can then be calculated as the sum over all trajectories with $K$ configuration changes, and the probability they occur,
\beq
    P_{t}(K) = \sum_{\omega_{t}} \pi(\omega_{t})\delta\big[\hat{K}(\omega_{t}) - K\big] \text{,}
    \label{Pk}
\eeq
where $\pi(\omega_{t})$ is the probability of observing $\omega_{t}$. For large times, this obeys the large deviation (LD) principle \cite{Touchette2009, Garrahan2007, Lecomte2007, Garrahan2009}
\beq
    P_{t}(K) \sim e^{t\varphi(K/t)} \text{,}
    \label{PkLD}
\eeq
where $\varphi(K/t)$ is called the LD rate function and plays the role of entropy density for trajectories. Alternatively, one can consider the moment generating function (MGF) \cite{Touchette2009} 
\beq
    Z_{t}(s) = \sum_{K} P_{t}(K)e^{-sK} = \sum_{\omega_{t}} \pi(\omega_{t})e^{-s\hat{K}(\omega_{t})} \text{,}
    \label{MGF}
\eeq
which contains equivalent information to \er{PkLD} and can be considered the partition function. From \er{MGF}, we see that the weighting of each trajectory is the probability that the trajectory occurs, exponentially re-weighted by its dynamical activity.
The MGF also obeys a LD principle,
\beq
    Z_{t}(s) \sim e^{t\theta(s)}\text{,}
    \label{MGFLD}
\eeq
where $\theta(s)$ is the scaled cumulant generating function (SCGF), whose derivatives evaluated at $s=0$ give the cumulants of $K$ scaled by time.
The SCGF plays the role of the thermodynamical free energy of trajectories and is related to the LD rate function by a Legendre transform $\theta(s) = -\min_{k} (sk + \varphi(k))$ \cite{Touchette2009}.

The MGF \er{MGF} can be expressed as 
\beq
    Z_{t}(s) = \braket{- | e^{t\W_{s}} | \text{in}} \text{,}
    \label{transfer}
\eeq
where $\ket{\text{in}}$ is some initial probability vector and $\W_{s}$ is a new operator which we name the {\em tilted generator}, and is a deformed version of \er{W} where we tilt with respect to the dynamical observable of interest \cite{Touchette2009, Garrahan2007, Lecomte2007, Garrahan2009}.
For the case of the dynamical activity \cite{Garrahan2007, Lecomte2007, Garrahan2009}, we simply tilt the off-diagonals of $\W$ with the same factor to obtain
\beq
    \W_{s} = \sum_{x, x'\neq x} e^{-s}w_{x\to x'}\ket{x'}\bra{x} - \sum_{x} R_{x}\ket{x}\bra{x} \text{.}
    \label{Ws}
\eeq
The largest eigenvalue of $\W_{s}$ is the SCGF $\theta(s)$, with associated left and right eigenvectors $\bra{l_{s}}$ and $\ket{r_{s}}$. Since $\bra{l_{s}}$ in general is not the flat state, $\W_{s}$ is not a proper stochastic generator for $s\neq 0$ \cite{Garrahan2007, Lecomte2007, Garrahan2009}. If one could exactly diagonalise \er{Ws} to find its leading eigenvalue and eigenvectors, then they would entirely unravel the LD statistics. We now briefly recap how this can be achieved using numerical TN techniques \cite{Banuls2019, Helms2019, Helms2020, Causer2020}.

\subsection{Variational matrix product states}
A matrix product state (MPS) is an ansatz for describing vector states of many-body systems \cite{Fannes1992,Oestlund1995,Vidal2003, Verstraete2004, Schollwoeck2011},  %orus2019
\beq
    \ket{\Psi} = \sum_{i_{1}, \dots, i_{N}}^{d} \Tr\big( A_{1}^{i_{1}}A_{2}^{i_{2}} \dots A_{N}^{i_{N}} \big) \ket{i_{1} \, i_{2}\, \dots \, i_{N}} \text{,}
    \label{MPS}
\eeq
where each subsystem $k$ has its own rank-3 tensor $A_{k}$ with the dimensions $d\times D \times D$.
The allowed entanglement within the state is controlled by the {\em bond dimension} $D$ \cite{Schuch2008}. 
It is often convenient to represent tensor networks in a diagrammatic form using shapes to represent tensorial objects, and (connecting) lines to represent contractions over tensors. For example, the corresponding diagram for an MPS is
\beq
\begin{tikzpicture}
    \node (text) at (0, 0) {$\ket{\Psi} = $};
    \node[draw, shape=circle, fill=black] (u0) at (0.8, -0.25) {};
    \node[draw, shape=circle, fill=black] (u1) at (1.55, -0.25) {};
    \node[draw, shape=circle, fill=black] (u2) at (2.3, -0.25) {};
    \node[draw, shape=circle, fill=black] (u3) at (3.05, -0.25) {};
    \draw[thick] (u0) -- (u1);
    \draw[thick] (u1) -- (u2);
    \draw[thick] (u2) -- (u3);
    \draw[thick] (u0) -- (0.8, 0.25);
    \draw[thick] (u1) -- (1.55, 0.25);
    \draw[thick] (u2) -- (2.3, 0.25);
    \draw[thick] (u3) -- (3.05, 0.25);
    \node (text3) at (3.5, 0) {,};
\end{tikzpicture}
\eeq
where each circle corresponds to one of the tensors $A_{k}$. Similarly, one can also attempt to write some operator $\hat{O}$ as a matrix product operator (MPO) \cite{Verstraete2004b,Zwolak2004,McCulloch2007, pirvu2010, Hubig2017, Parker2020}.
Operators which act locally on the sub-systems, such as \ers{Wkmc}{WSSEP}, can be efficiently described as a MPO. That is to say we can represent them exactly in MPO form with only a small constant bond dimension. The diagrammatic representation for MPOs is
\beq
\begin{tikzpicture}
    \node (text) at (0, 0) {$\hat{O} = $};
    \node[draw, shape=rectangle, fill=black] (u0) at (0.8, 0) {};
    \node[draw, shape=rectangle, fill=black] (u1) at (1.55, 0) {};
    \node[draw, shape=rectangle, fill=black] (u2) at (2.3, 0) {};
    \node[draw, shape=rectangle, fill=black] (u3) at (3.05, 0) {};
    \draw[thick] (u0) -- (u1);
    \draw[thick] (u1) -- (u2);
    \draw[thick] (u2) -- (u3);
    \draw[thick] (u0) -- (0.8, -0.5);
    \draw[thick] (u1) -- (1.55, -0.5);
    \draw[thick] (u2) -- (2.3, -0.5);
    \draw[thick] (u3) -- (3.05, -0.5);
    \draw[thick] (u0) -- (0.8, 0.5);
    \draw[thick] (u1) -- (1.55, 0.5);
    \draw[thick] (u2) -- (2.3, 0.5);
    \draw[thick] (u3) -- (3.05, 0.5);
    \node (text3) at (3.5, 0) {.};
\end{tikzpicture}
\eeq

MPS allow for the easy and efficient implementation of the widely used density matrix renormalization group (DMRG) method \cite{White1992,Schollwoeck2005}, an algorithm designed to iteratively minimize the energy of a state $E_{\Psi}$ with respect to some Hamiltonian $\hat{H}$. 
In the language of MPS \cite{Schollwoeck2011}, we start with some guess at some fixed bond dimension, and sweep through each tensor applying local optimizations with all other tensors fixed. This is done until we reach convergence, which is usually when the change in energy of the state per sweep is small.
At the end of the routine, one can efficiently calculate the variance of the state with respect to the Hamiltonian
\beq
    \delta {E_{\Psi}}^{2} = \text{var}_{\hat{H}}(\Psi) = \braket{\hat{H}^{2}}_{\Psi} - \braket{\hat{H}}^{2}_{\Psi}
    \label{MPS_variance}
\eeq
where $\braket{\cdot}_{\Psi} = \braket{\Psi | \cdot | \Psi}$ denotes an expectation value.
We check to see if it has fallen below some desired value, $\epsilon$; if not, we run the algorithm with an increased bond dimension, where we typically use the state from the previous run as an initial guess. For more details on the workings of variational MPS (vMPS) algorithms, see the reviews \cite{Schollwoeck2011, Verstraete2008}.

Many recent works have shown that vMPS algorithms are very effective for studying the LD statistics of classically constrained systems which obey detailed balance \cite{Banuls2019, Causer2020, Helms2019}.
In particular, if we write the tilted generator in a way such that it is Hermitian then the state we are searching for is the ground state. This guarantees each update is an improvement upon the last.
For dynamics obeying detailed balance, the activity-tilted generator can be brought to a Hermitian form using a similarity transformation that is independent of $s$ \cite{Garrahan2009},
\beq
    \mathbb{H}_{s} = -Q^{-1}\W_{s}Q . 
    \label{similarity_transformation}
\eeq
For the case of the East/FA models \cite{Garrahan2009}, the diagonal operator $Q$ is given by
\beq
    Q^{\text{FA/East}} = \big[ \sqrt{1-c}\ket{0}\bra{0} + \sqrt{c}\ket{1}\bra{1} \big]^{\otimes N} , 
\eeq
and for the SSEP by $Q^{\text{SSEP}} = \mathbb{I}$. The new Hamiltonian $\mathbb{H}_{s}$ has the ground state $\ket{\psi_{s}}$ with energy $-\theta(s)$. The ground state is related to the left and right eigenvectors of $\mathbb{W}_{s}$ in the following way \cite{Banuls2019},
\begin{align}
    \ket{\psi_{s}} &= Q^{-1} \ket{r_{s}} ,
    \label{psi_right}
    \\
    \bra{\psi_{s}} &= \bra{l_{s}}Q ,
    \label{psi_left}
    \\
    \ket{\psi_{s}} &= \sum_{x} \sqrt{l_{s}(x)r_{s}(x)}\ket{x} ,
    \label{psi_definition}
\end{align}
where $l_{s}(x) = \braket{l_{s} | x}$ and  $r_{s}(x) = \braket{x | r_{s}}$.

\section{Doob transformation and Optimal sampling}
\label{sec_doob}
We now define the so-called generalised Doob transformation \cite{Simon2009, Popkov2010, Jack2010, Chetrite2015, Carollo2018, Oakes2018, oakes2020}, and show how one can use our MPS solution to \er{similarity_transformation} to construct a reference dynamics which closely resembles the true Doob dynamics. We then present a method to optimally sample the rare events of our toy models using these new dynamics.

\subsection{Generalised Doob dynamics}
The goal is to find a proper stochastic generator which generates trajectories with the same probabilities as those in the tilted dynamics $\W_{s}$, cf.~\er{Ws}. This can be achieved using the (long-time) generalised Doob transformation \cite{Simon2009, Popkov2010, Jack2010, Chetrite2015, Carollo2018, Oakes2018, oakes2020}, defined as 
\beq
    \W_{s}^{\text{Doob}} = \mathbb{L}[\W_{s} - \theta(s)\mathbb{I}]\mathbb{L}^{-1} \text{,}
    \label{Wdoob}
\eeq
where $\mathbb{L} = \text{diag}(\bra{l_{s}})$ is the left eigenvector $\bra{l_{s}}$ as a diagonal matrix.
It is easy to check that \er{Wdoob} is annihilated by the flat state $\bra{-}$, which means that 
$\W_{s}^{\text{Doob}}$ is a stochastic operator. Its stationary state is 
\beq
    \ket{\text{ss}}^{\text{Doob}}_{s} = \sum_{x} \tilde{P}(x) \ket{x} = \sum_{x} l_{s}(x)r_{s}(x) \ket{x}.
\eeq
The generator $\W^{\text{Doob}}_{s}$ can also be expressed as a sum of its diagonal and off-diagonal elements
\begin{align}
    \W^{\text{Doob}}_{s} = \sum_{x, x'\neq x} & \frac{l_{s}(x')}{l_{s}(x)}e^{-s} w_{x\to x'}\ket{x'}\bra{x}
    \nonumber
    \\
    & - \sum_{x} (R_{x} + \theta(s))\ket{x}\bra{x} \text{.}
\end{align}
Thus our new dynamics has the transition rates and escape rates
\beq
    \tilde{w}_{x\to x'} = \frac{l_{s}(x')}{l_{s}(x)}e^{-s} w_{x\to x'}
    \label{doob_flips}
\eeq
\beq
    \tilde{R}_{x} = R_{x} + \theta(s)
    \label{doob_escapes}
\eeq
respectively. That is to say the transition rates are re-weighted by $e^{-s}$ and by some ratio ${l_{s}(x')}/{l_{s}(x)}$ which depends on the structure of the configurations, and the escape rate is shifted by $\theta(s)$.

We now consider some general time-dependent observable $\hat{A}$, and ask what is the expectation value in the tilted dynamics,
\beq
    \braket{\hat{A}}_{s} \equiv \frac{\braket{\hat{A}e^{-sK}}}{\braket{e^{-sK}}} = Z_{t}(s)^{-1}\sum_{\omega_{t}} \pi(\omega_{t})\hat{A}(\omega_{t})e^{-s\hat{K}(\omega_{t})} \text{.}
    \label{expectation_s}
\eeq
One can now apply importance sampling to arrive at
\begin{align}
    \braket{\hat{A}}_{s} &= Z_{t}(s)^{-1}\sum_{\omega_{t}} \tilde{\pi}(\omega_{t}) \frac{\pi(\omega_{t})}{\tilde{\pi}(\omega_{t})} \hat{A}(\omega_{t})e^{-s\hat{K}(\omega_{t})} \text{,}
    \nonumber
    \\
    &=  Z_{t}(s)^{-1} \bigg<\frac{\pi}{\tilde{\pi}} \hat{A} e^{-s\hat{K}}\bigg>_{\text{Doob}}
    \label{expectation_doob}
\end{align}
where $\tilde{\pi}(\omega_{t})$ is the probability of observing $\omega_{t}$ in the dynamics generated by $\W_{s}^{\text{Doob}}$ and $\braket{\bf{\cdot}}_{\text{Doob}}$ denotes an expectation value with respect to trajectories with probabilities from the Doob dynamics. At a first glance, it might look that we have  not gained much from expressing the expectation of $A$ using the Doob generator $\W_{s}^{\text{Doob}}$. However, if one calculates the ratio of probabilities in \er{expectation_doob} then the power of this expression becomes apparent.

Let us first consider the original dynamics described by \er{W}. If we have some system in configuration $x$, then the probability it flips to some other state $x'$ at the time $\Delta t$ is 
\beq
    P_{x\to x'}(\Delta t) = w_{x\to x'}e^{-R_{x}\Delta t} \text{.}
\eeq
It then follows that the trajectory $\omega_{t}$ occurs with probability 
\beq
\pi(\omega_{t}) = P(x_0) \, e^{-R_{x_K}(t-t_{x_K})}\prod_{i=1}^{K} w_{x_{i-1}\to x_{i}} \, e^{-R_{x_{i-1}}(t_{x_i}-t_{x_{i-1}})} ,
\eeq
where we have also accounted for the fact that the system must remain in the same state after the final flip for the remainder of the time and the probability of the initial configuration $P(x_0)$ (where we assume it is in the steady state). 
The probability of the trajectory under the Doob dynamics has a similar form, with the substitutions $w_{x\to x'} \to \tilde{w}_{x\to x'}$, $R_{x} \to \tilde{R}_{x}$, and $P(x_0) \to \tilde{P}(x_0)$,  
\begin{align}
    \tilde{\pi}(\omega_{t}) = & \tilde{P}(x_0) \, e^{-sK}e^{-t\theta(s)} \frac{l_{s}(x_{K})}{l_{s}(x_{0})} e^{-R_{x_{k}}(t-t_{k})}
    \\
    & \times \prod_{i=1}^{K} w_{x_{i-1}\to x_{i}} \, e^{-R_{x_{i-1}}(t_{x_{i}}-t_{x_{i-1}})} \text{,}
    \nonumber
\end{align}
where all but the endpoint factors of $l_{s}(x)$ cancel out telescopically. The ratio of probabilities then goes as
\beq
    \frac{\pi(\omega_{t})}{\tilde{\pi}(\omega_{t})} = 
    \frac{e^{sK}e^{t\theta(s)} }{l_{s}(x_{0}) \, l_{s}(x_{K})} \text{,}
    \label{prob_ratio}
\eeq
where we have used $\tilde{P}(x_{0}) = P(x_{0}) \, l_{s}(x_{0})^{2}$. Substituting \er{prob_ratio} back into \er{expectation_doob} cancels out the exponential tilting $e^{sK}$. Furthermore, for large times, $Z_{t}(s)^{-1} \approx e^{-t\theta(s)}$ giving the final result
\beq
    \braket{\hat{A}}_{s} = \bigg<
    \frac{1}{l_{s}(x_{0}) \, l_{s}(x_{K})}
	\hat{A}\bigg>_{\text{Doob}} \text{.}
	\label{AAD}
\eeq
And so it follows that one can exactly sample the expectation value of a trajectory observable in the tilted ensemble defined by the non-stochastic tilted generator, by sampling it directly from trajectories generated by the stochastic Doob dynamics \er{Wdoob}, up to factors at the endpoints of each trajectory (which become negligible in the long time limit if $\hat{A}$ is time-extensive). 
We note that \er{AAD} can also be derived by means of linear algebra, using \er{Wdoob} and the ratio $P(x_{0}) / \tilde{P}(x_{0})$ (see Ref.~\cite{Tizon-Escamilla2019} for details).

\subsection{Reference Dynamics}
While the above shows how to optimally sample if one has access to the Doob generator, which is obtained from the exact minimisation of the tilted generator, we now consider how to approximate it efficiently.  

Suppose we have an MPS approximation $\ket{\psi_{s}^{\text{ref}}}$ to the ground state of the Hermitian operator $\mathbb{H}_{s}$, where our choice of bond dimension $D$ controls the error. By applying the operator $Q^{-1}$ to $\ket{\psi_{s}^{\text{ref}}}$, as is done in \er{psi_left}, one can also retrieve an approximation to the left eigenvector. This is easily done as an MPS-MPO product,
\beq
\begin{tikzpicture}
    \node (text) at (-0.95, 0) {$\bra{l_{s}^{\text{ref}}} = \bra{\psi_{s}^{\text{ref}}}Q^{-1} = $};
    \node[draw, shape=circle, fill=black] (v0) at (1, 0.5) {};
    \node[draw, shape=circle, fill=black] (v1) at (1.75, 0.5) {};
    \node[draw, shape=circle, fill=black] (v2) at (2.5, 0.5) {};
    \node[draw, shape=circle, fill=black] (v3) at (3.25, 0.5) {};
    \node[draw, shape=rectangle, fill=black] (m0) at (1, 0) {};
    \node[draw, shape=rectangle, fill=black] (m1) at (1.75, 0) {};
    \node[draw, shape=rectangle, fill=black] (m2) at (2.5, 0) {};
    \node[draw, shape=rectangle, fill=black] (m3) at (3.25, 0) {};
    \draw[thick] (v0) -- (v1);
    \draw[thick] (v1) -- (v2);
    \draw[thick] (v2) -- (v3);
    \draw[thick] (v0) -- (m0);
    \draw[thick] (v1) -- (m1);
    \draw[thick] (v2) -- (m2);
    \draw[thick] (v3) -- (m3);
    \draw[thick] (m0) -- (1, -0.5);
    \draw[thick] (m1) -- (1.75, -0.5);
    \draw[thick] (m2) -- (2.5, -0.5);
    \draw[thick] (m3) -- (3.25, -0.5);
    \node (text2) at (0.4, -1.25) {=};
    \node[draw, shape=circle, fill=black] (u0) at (1, -1) {};
    \node[draw, shape=circle, fill=black] (u1) at (1.75, -1) {};
    \node[draw, shape=circle, fill=black] (u2) at (2.5, -1) {};
    \node[draw, shape=circle, fill=black] (u3) at (3.25, -1) {};
    \draw[thick] (u0) -- (u1);
    \draw[thick] (u1) -- (u2);
    \draw[thick] (u2) -- (u3);
    \draw[thick] (u0) -- (1, -1.5);
    \draw[thick] (u1) -- (1.75, -1.5);
    \draw[thick] (u2) -- (2.5, -1.5);
    \draw[thick] (u3) -- (3.25, -1.5);
    \node (text3) at (3.75, -1.25) {.};
\end{tikzpicture}
\eeq
We then construct the generator of the so-called {\em reference dynamics}, which goes as \er{W} with the transition rates and escape rates given by
\begin{align}
    w^{\rm ref}_{x\to x'} &= \frac{l_{s}^{\rm ref}(x')}{l_{s}^{\rm ref}(x)}e^{-s} w_{x\to x'} ,
    \label{ref_flips}
    \\
    {R}^{\rm ref}_{x} &= \sum_{x'\neq x} w^{\rm ref}_{x\to x'},
    \label{ref_escapes}
\end{align}
respectively. Note that here we have not used \er{doob_escapes} for the escape rates, as these reference dynamics only act as an approximation to the Doob dynamics, and thus would not give a true stochastic dynamics.
In appendix A, we show the steady-state solution to the reference dynamics is given by
\beq
    \ket{\text{ss}}^{\text{ref}}_{s} = \sum_{x} \psi^{\text{ref}}_{s}(x)^{2} \ket{x},
\eeq
where $\psi^{\text{ref}}_{s}(x) = \braket{x | \psi^{\text{ref}}_{s}}$.

If we repeat the steps between \era{expectation_doob}{AAD} but for the reference dynamics, the expectation \er{expectation_doob} looks like
\beq
    \braket{\hat{A}}_{s} = \bigg< 
    \frac{1}{l_{s}^{\text{ref}}(x_{0}) \, l_{s}^{\text{ref}}(x_{K})}
    e^{-t\theta(s) + \int dt \Delta \hat{R}}  \hat{A}\bigg>_{\text{ref}},
    \label{doob_umbrella}
\eeq
where 
$\int dt \Delta \hat{R}$ is the time integral of the difference of escape rates between the reference dynamics and the original dynamics, with $\Delta \hat{R}_x = {R}^{\rm ref}_{x} - R_x$. 
We can estimate a sampling error when using \er{doob_umbrella} in the following way \cite{Oakes2018}. First, let us assume the effects of the time-edge factors is negligible (as they are not exponential in time) and try to sample the quantity
\beq
    \braket{e^{-s\hat{K}}} = \braket{\mathcal{R}e^{-s\hat{K}}}_{\text{ref}} \approx \frac{1}{N_{\text{sp}}}\sum_{\alpha=1}^{N_{\text{sp}}}  \mathcal{R}(\omega^{\alpha})e^{-s\hat{K}(\omega^{\alpha})},
    \label{sampling}
\eeq
where $\mathcal{R}(\omega^{\alpha}) = e^{s\hat{K}(\omega^{\alpha}) + \int dt \Delta \hat{R}(\omega^{\alpha})}$ is the umbrella which compensates for change in sampling dynamics and we estimate for a fixed number of samples, $N_{\text{sp}}$.
The variance of \er{sampling} gives a way to quantify the sampling error,
\begin{align}
    \epsilon_{\text{ref}}^{2} &= \frac{\Var_{\text{ref}}\left( \frac{1}{N_{\text{sp}}}\sum_{\alpha=1}^{N_{\text{sp}}}  \mathcal{R}(\omega^{\alpha})e^{-s\hat{K}(\omega^{\alpha})} \right)}
    {\left<\frac{1}{N_{\text{sp}}}\sum_{\alpha=1}^{N_{\text{sp}}}  \mathcal{R}(\omega^{\alpha})e^{-s\hat{K}(\omega^{\alpha})}\right>^{2}_{\text{ref}}}
    \nonumber
    \\
    &= \frac{1}{N_{\text{sp}}} \left[ \frac{\braket{\mathcal{R}^{2}e^{-2s\hat{K}}}_{\text{ref}}} {\braket{\mathcal{R}e^{-s\hat{K}}}_{\text{ref}}^{2}} - 1 \right]
    \nonumber
    \\
    &= \frac{1}{N_{\text{sp}}} \left[ \frac{\braket{e^{2\int dt' \Delta\hat{R}}}_{\text{ref}}} {\braket{e^{\int dt' \Delta\hat{R}}}_{\text{ref}}^{2}} - 1 \right].
    \label{sampling_error_ER}
\end{align}
In appendix B we show
\beq
    \epsilon_{\text{ref}}^{2} \approx \frac{e^{t \delta{E}^{2}} - 1}{N_{\text{sp}}}
    \approx \frac{t \delta E^{2}}{N_{\text{sp}}} .
    \label{approx}
\eeq
The last approximation holds for $\delta{E}$ small enough ($t \delta{E}^{2} \ll 1$). In \er{approx}, $\delta E^{2}$ is the calculated variance on our MPS approximation of the leading eigenvector, see \er{MPS_variance}.

\subsection{Simulating trajectories}
We are now in  a position to efficiently simulate trajectories from our reference dynamics.
The sampling of trajectories from a classical generator is usually achieved using a continuous time Monte Carlo (CTMC, otherwise known as the BKL algorithm) \cite{Bortz1975}.
Given that our system is in some configuration $x$ at time $t'$, we need to calculate the next jump in the trajectory. That is, we need to decide the next configuration the system will move into, and the time it does so.
Calculating this can be split into five separate steps:
\begin{enumerate}
\item Find each configuration $x'$ the system can move into from $x$.
\item Calculate the transition rates $w_{x\to x'}$ for each $x'$.
\item Calculate the escape rate $R_{x}$ as the sum of all transition rates.
\item Randomly choose one $x'$, each with the probability $w_{x\to x'} / R_{x}$
\item Randomly choose the jump time $\Delta t$ with probability
$P(\Delta t) = R_{x} e^{-R_{x} \Delta t}$.
\end{enumerate}
By starting at a configuration sampled from equilibrium (which in the case of the reference dynamics can be efficiently done using the MPS $\ket{\psi^{\text{ref}}_{s}}$ \cite{Ferris2012,Iblisdir2014}), or otherwise, one can simply repeat this procedure until some total time $t$ has elapsed.

We can use this method for our reference dynamics, where the only step that needs slight adjustment is the second.
While one must still calculate the transition rates of the original dynamics in the usual way, we must also calculate the left vector components $l_{s}^{\text{ref}}(x)$ and $l_{s}^{\text{ref}}(x')$.
Let us assume the former is carried over from the previous jump in the algorithm. Then all one needs to do is calculate each $l_{s}^{\text{ref}}(x')$. We start by noting that any configuration $x$ can be written in MPS form with bond dimension $1$,
\beq
\begin{tikzpicture}
    \node (text) at (0, 0) {$\ket{x} = \ket{n_{1}}\otimes\ket{n_{2}}\otimes\dots\otimes\ket{n_{N}}$};
    \node (text2) at (-1.6, -0.75) {$=$};
    \node[draw, shape=circle] (u0) at (-1.1, -1) {};
    \node[draw, shape=circle] (u1) at (-0.35, -1) {};
    \node[draw, shape=circle] (u2) at (0.4, -1) {};
    \node[draw, shape=circle] (u3) at (1.15, -1) {};
    \draw[thick] (u0) -- (-1.1, -0.5);
    \draw[thick] (u1) -- (-0.35, -0.5);
    \draw[thick] (u2) -- (0.4, -0.5);
    \draw[thick] (u3) -- (1.15, -0.5);
\end{tikzpicture}
\eeq
and then we can simply calculate the left component as a MPS-MPS contraction
\beq
\begin{tikzpicture}
    \node (text) at (-0.8, 0) {$l_{s}^{\text{ref}}(x) \, = \, |\braket{l_{s}^{\text{ref}} | x}| = $};
    \node[draw, shape=circle, fill=black] (v0) at (1, 0.25) {};
    \node[draw, shape=circle, fill=black] (v1) at (1.75, 0.25) {};
    \node[draw, shape=circle, fill=black] (v2) at (2.5, 0.25) {};
    \node[draw, shape=circle, fill=black] (v3) at (3.25, 0.25) {};
    \node[draw, shape=circle] (u0) at (1, -0.25) {};
    \node[draw, shape=circle] (u1) at (1.75, -0.25) {};
    \node[draw, shape=circle] (u2) at (2.5, -0.25) {};
    \node[draw, shape=circle] (u3) at (3.25, -0.25) {};
    \draw[thick] (v0) -- (v1);
    \draw[thick] (v1) -- (v2);
    \draw[thick] (v2) -- (v3);
    \draw[thick] (v0) -- (u0);
    \draw[thick] (v1) -- (u1);
    \draw[thick] (v2) -- (u2);
    \draw[thick] (v3) -- (u3);
    \node (text2) at (3.5, 0) {.};
\end{tikzpicture}
\eeq
The transition rates for the reference dynamics are then calculated using \er{ref_flips}, and the method proceeds as before.
The total computational cost for calculating each $l_{s}^{\text{ref}}(x)$ is $O(D^{2}N)$, and thus the total cost of each Monte Carlo (MC) step is $O(D^{2}NN_{F})$, where $N_{F}$ is the total number of configurations $x'$ for a given step.

\begin{figure*}[t]
    \centering
    \includegraphics[width=\linewidth]{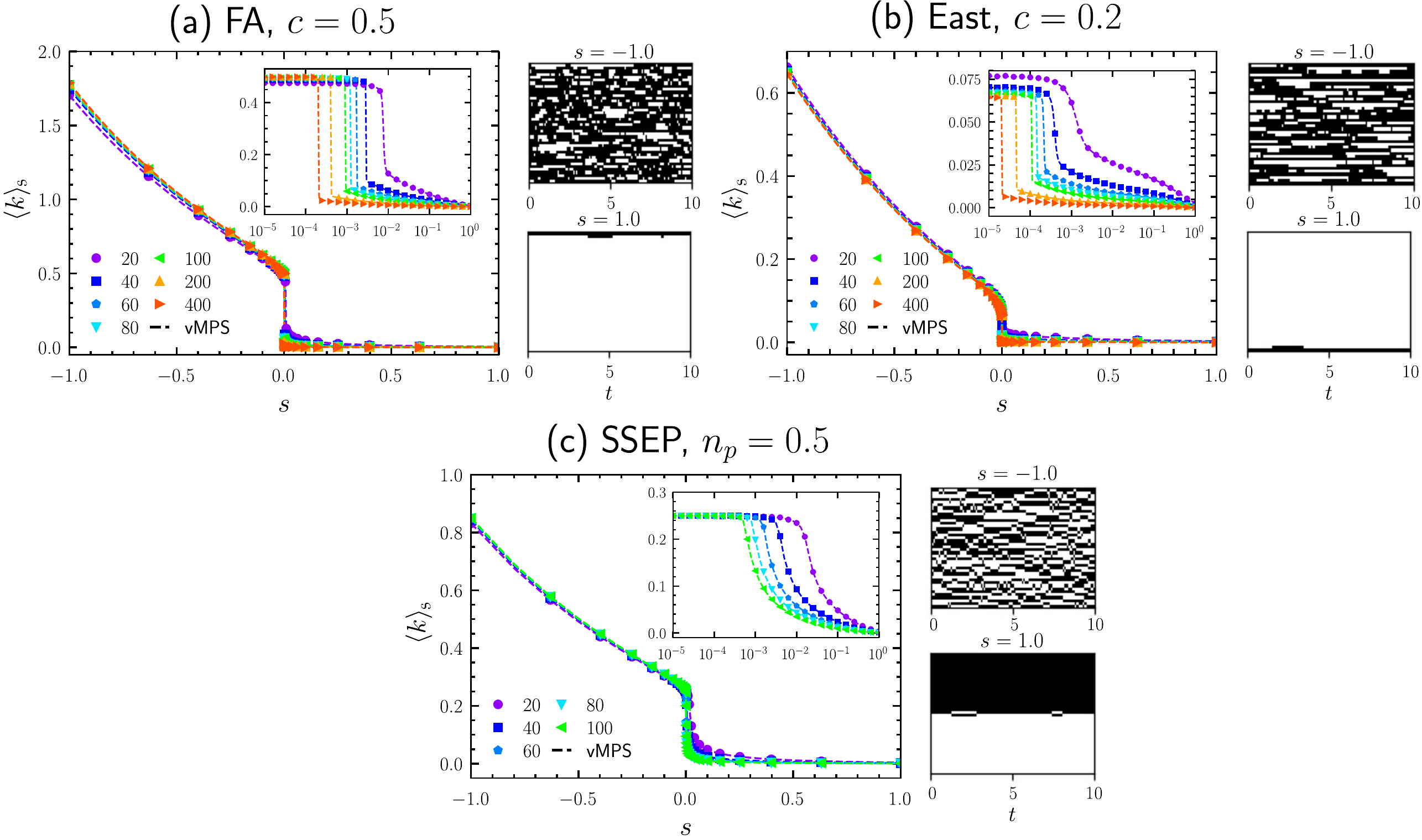}
    \caption{\textbf{The dynamical activity from brute-force Monte Carlo.} We show the dynamical activity measured for (a) the FA model with $c = 0.5$, (b) the East model with $c = 0.2$ and (c) SSEP with $n_{p} = 0.5$. A variety of system sizes $N\in[20, 400]$ are shown for each. The dashed lines show expected activity calculated directly from the MPS $\Tilde{k}(s)$, whereas the markers show the activity measured via CTMC $\braket{k}_{s}$ with a time $t = 100 / \Tilde{k}(s)$. The inactive phase is shown with a log $s$ scale in the insets. We also show representative trajectories at $s = -1, 1$ for each model.
    }
    \label{fig: doob_brute_force}
\end{figure*}

Let us now consider our KCMs where we have single-spin flip dynamics. We first note that the number of possible configuration changes from $x$ is bounded by the number of sites, that is, $1\leq N_{F} \leq N$. Using the method described above, the computational cost for each step is at worst, quadratic in the system size. However, by realising that the tensor network contractions $\braket{l_{s}^{\text{ref}} | x}$ and $\braket{l_{s}^{\text{ref}} | x'}$ are identical apart from just one tensor (corresponding to the spin which would flip), we can reduce the computational cost by recycling partial contractions from the edges. We first need to identify the first and last sites on the lattice which are able to flip, which we label $i_{l}$ and $i_{r}$ respectively. In a similar fashion to variational algorithms, we then contract from the left edge of tensor network $\braket{l_{s}^{\text{ref}} | x}$ up to $i_{r}-1$, and saving each tensor block along the way.
\beq
\begin{tikzpicture}
    \node[draw, shape=circle, fill=black] (v0) at (1, 0.25) {};
    \node[draw, shape=circle, fill=black] (v1) at (1.75, 0.25) {};
    \node[draw, shape=circle, fill=black] (v2) at (2.5, 0.25) {};
    \node[draw, shape=circle, fill=black] (v3) at (3.25, 0.25) {};
    \node[draw, shape=circle] (u0) at (1, -0.25) {};
    \node[draw, shape=circle] (u1) at (1.75, -0.25) {};
    \node[draw, shape=circle] (u2) at (2.5, -0.25) {};
    \node[draw, shape=circle] (u3) at (3.25, -0.25) {};
    \draw[thick] (v0) -- (v1);
    \draw[thick] (v1) -- (v2);
    \draw[thick] (v2) -- (v3);
    \draw[thick] (v0) -- (u0);
    \draw[thick] (v1) -- (u1);
    \draw[thick] (v2) -- (u2);
    \draw[thick] (v3) -- (u3);
    
    \node (text2) at (3.75, 0) {$\rightarrow$};
    
    \node[draw, shape=rectangle, minimum width = 0.4cm, minimum height = 1cm, fill=black] (b1) at (4.5, 0) {};
    \node[draw, shape=circle, fill=black] (v21) at (5.25, 0.25) {};
    \node[draw, shape=circle, fill=black] (v22) at (6, 0.25) {};
    \node[draw, shape=circle, fill=black] (v23) at (6.75, 0.25) {};
    \node[draw, shape=circle] (u21) at (5.25, -0.25) {};
    \node[draw, shape=circle] (u22) at (6, -0.25) {};
    \node[draw, shape=circle] (u23) at (6.75, -0.25) {};
    \draw[thick] (4.5, 0.25) -- (v21);
    \draw[thick] (v21) -- (v22);
    \draw[thick] (v22) -- (v23);
    \draw[thick] (v21) -- (u21);
    \draw[thick] (v22) -- (u22);
    \draw[thick] (v23) -- (u23);
    
    \node (text2) at (3.75, -1.25) {$\rightarrow$};
    
    \node[draw, shape=rectangle, minimum width = 1.2cm, minimum height = 1cm, fill=black] (b1) at (4.9, -1.25) {};
    \node[draw, shape=circle, fill=black] (v32) at (6, -1) {};
    \node[draw, shape=circle, fill=black] (v33) at (6.75, -1) {};
    \node[draw, shape=circle] (u32) at (6, -1.5) {};
    \node[draw, shape=circle] (u33) at (6.75, -1.5) {};
    \draw[thick] (4.5, -1) -- (v32);
    \draw[thick] (v32) -- (v33);
    \draw[thick] (v32) -- (u32);
    \draw[thick] (v33) -- (u33);
\end{tikzpicture}
\nonumber
\eeq
We do the same but from the right and up to $i_{l}+1$.
This initialization of partial contractions has a one-time cost of 
\beq
    O(D^{2}(N +i_{r} - i_{l} - 2)) < O(2D^{2}N).
\eeq

Calculating each $l_{s}^{\text{ref}}(x')$ at site $j$ is then easy.
We just contract our remaining tensors at site $j$ with the previously saved left and right blocks,
\beq
\begin{tikzpicture}
    \node (text) at (-1.2, 0) {$l_{s}^{\text{ref}}(x') = $};
    \node[draw, shape=rectangle, minimum width = 0.4cm, minimum height = 1cm, fill=black] (b1) at (0, 0) {};
    \node[draw, shape=circle, fill=black] (v) at (0.75, 0.25) {};
    \node[draw, shape=circle] (u) at (0.75, -0.25) {};
    \node[draw, shape=rectangle, minimum width = 0.4cm, minimum height = 1cm, fill=black] (b2) at (1.5, 0) {};
    \draw[thick] (0, 0.25) -- (v);
    \draw[thick] (1.5, 0.25) -- (v);
    \draw[thick] (u) -- (v);
    \node (text) at (2, 0) {.};
\end{tikzpicture}
\nonumber
\eeq
This is done for each possible site which can flip, and thus entails a computational cost $O(D^{2}N_{F})$.
Once a choice is made for which site to flip, which we will label $i$, we must update the blocks of partial contractions up to (the now possibly different) $i_{l}$ and $i_{r}$.
Note that this time we do not have to start from the edges of the MPS, but just from site $i$ as the previous partial contractions that come before do not change.
The total cost of updating the partial contractions is 
\beq
    O\big(D^{2}[(i_{r}-i) + (i - i_{l})]\big) = O\big( D^{2} (i_{r} - i_{l} ) \big) \text{.}
\eeq
The total computation cost for each MC step is the sum of the cost for calculating each $l_{s}^{\text{ref}}(x)$ and updating the partial blocks after a choice is made,
\beq
    O\big( D^{2}(N_{F} + i_{r} - i_{l}) \big) \leq O\big( 2D^{2}N \big).
\eeq
Consequentially, the cost of each MC step is reduced to one which is at most linear in system size.

\section{Numerical Results}
\begin{figure*}[t]
    \centering
    \includegraphics[width=\linewidth]{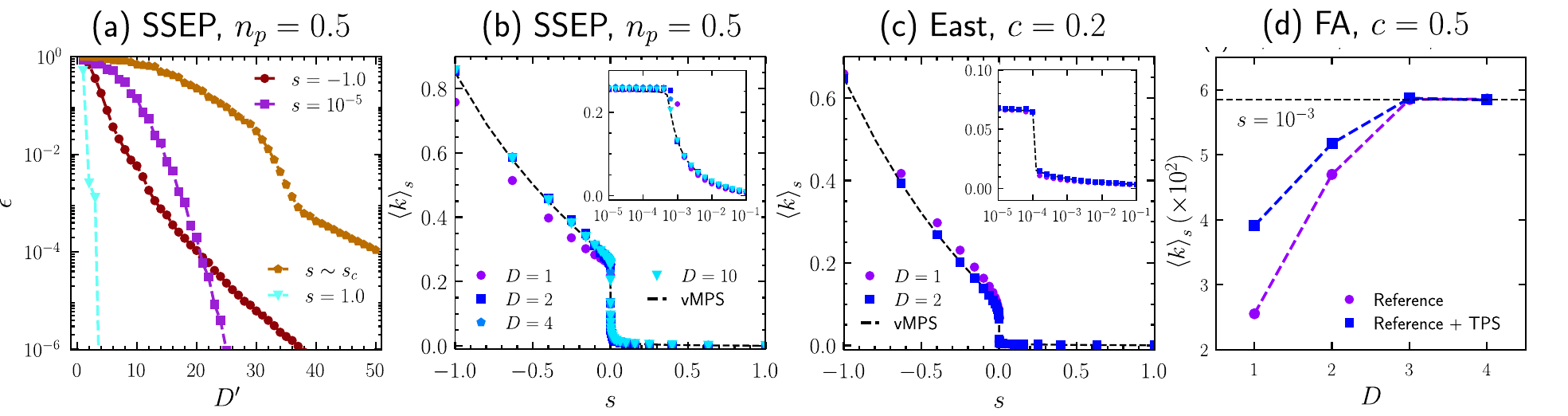}
    \caption{\textbf{Reference dynamics from a truncated MPS.} 
    All data is for the system size $N = 100$.
    (a) The truncation error $\varepsilon = 1 - |\braket{\psi_{D} | \psi_{D'}}|^{2}$ as a function of the truncated bond dimension $D'$ for SSEP at various values of $s$.
    (b) The measured average dynamical activity $\braket{k}_{s}$ with a reference dynamics constructed with truncated MPS. The dashed line shows the expected value obtained through vMPS with a large ($D\geq50$) bond dimension, and the inset shows the same but on a log-scale around the critical point.
    (c) The same but for the East model.
    (d) The measured dynamical activity $\braket{k}_{s}$ as a function of $D$ for $s = 10^{-3}$ close to the critical point $s_{c}$. The purple circles show the values measured using the reference dynamics alone, whereas the blue squares show values obtained using the reference dynamics and TPS to incorporate umbrella sampling. The dashed line shows the expected value obtained from vMPS. Each point is done for a trajectory time of $t = 100$ and $N_{sp}=10^{6}$ trajectories.
    }
    \label{fig: truncated}
\end{figure*}
\subsection{Approximating the Doob Dynamics}

We put to the test the general method presented above by approximating the Doob dynamics of each model defined in Sec.~II. We show that the Doob dynamics is well estimated using the MPS reference dynamics, and can even be well approximated with truncated MPS.

Each of the three models is known to exhibit a trajectory phase transition (when tilted against the activity) for long times and 
in the thermodynamic limit $N\to\infty$, manifested in the SCGF $\theta(s)$ at $s=0$ with a discontinuous drop in the dynamical activity $\hat{K}(s) = -\theta'(s)/N$ \cite{Garrahan2007,Appert-Rolland2008,Garrahan2009,Jack2015,Causer2020}.
We call the dynamical phase for $s<0$ the {\em active phase}, and that for $s>0$, the {\em inactive phase}.
One is able to do a detailed investigation of this first-order phase transition by considering the finite-size scaling of the model \cite{Bodineau2012, Bodineau2012b, Nemoto2017, Banuls2019, Causer2020}.
We can estimate a critical point $s_{c}(N) \gtrsim 0 $ by finding the peak of the dynamical susceptibility $\chi(s) = \theta''(s)$, which shows a drastic change in a small region around the transition point. 

We start by taking the usual approach of approximating the ground states $\ket{\psi_{s}}$ using vMPS.  That is, we run the algorithm allowing the bond dimension to increase until the variance of the energy (with respect to the Hamiltonian) falls sufficiently, cf.~\er{MPS_variance}.
The resultant MPS is then used to construct the reference dynamics, which approximates the Doob dynamics to a high accuracy, as explained in the previous section. 
Note that because the vMPS tries to keep entanglement as low as possible, for $s>s_c(N)$ the approximated ground state exhibits localisation at just one edge of the system \cite{Banuls2019}.
While for the East case this corresponds to the structure of the ground state in the sector with fixed occupation 1 in the leftmost site, the FA and SSEP models
have reflection symmetry, spontaneously broken for $s>0$ and large $N$.
Thus, in order to maintain the symmetry in the latter two cases, we construct an MPS which is a superposition of the result from vMPS and its spatially reflected state to obtain our dynamics in the inactive phase.

\subsubsection{Direct sampling with the reference dynamics but without re-weighting}

We first check that the CTMC algorithm with our MPS reference dynamics gives the expected results. We do this {\em  without using} the trajectory re-weighting, cf.~\er{doob_umbrella}. This amounts to only considering infinite-time dynamics, and assuming that our approximation is actually exact. Despite this strong assumption, we find that it produces excellent results as shown in Fig.~\ref{fig: doob_brute_force}.
The expected dynamical activity (per unit site and time, dashed lines) can be calculated as a TN contraction over our MPS and MPO,
\beq
    \Tilde{k}(s) = \frac{1}{N} \Big<\psi_{s} \Big | \frac{d  \mathbb{H}_{s}}{ds} \Big | \psi_{s} \Big> .
    \label{activity_MPS}
\eeq
The same quantity can be calculated on a trajectory level (symbols) by counting the total number of configuration changes, $\braket{K}$ and taking its time (and spatial) average,
\beq
    \braket{k}_{s} = \frac{\braket{K}}{Nt} ,
    \label{activity_MC}
\eeq
where $t$ is the run time for each trajectory.
We show results for each model, for a range of system sizes of $N\in[20, 400]$. The expected and measured results have excellent agreement.
This simplified algorithm struggles most around the transition point, $s_{c}(N)$, due to the required large bond dimension (see Refs.~\cite{Banuls2019, Causer2020}).

We also show representative trajectories for the active ($s = -1$) and inactive phases ($s = 1$). Each model excellently demonstrates the difference in dynamics between the two phases.
The active phase displays very rapid changes with structures that allow for unconstrained dynamics. For the FA and East models this means having a large number of excitations, while SSEP requires particles to be spaced apart.
Conversely, this inactive phase has just few configuration changes with highly constrained dynamics. This means minimizing the number of excitations for the FA and East resulting in the dynamics responsible for the so-called ``space-time bubble'' in local regions of space \cite{Garrahan2002b, Chandler2010, Garrahan2009}, while for SSEP we restrict the activity by clustering the particles \cite{Lecomte2012,Jack2015}. To our knowledge, direct dynamical sampling of trajectories for these systems sizes and values of $s \neq 0$ is unprecedented for these three models.

\begin{figure*}[t]
    \centering
    \includegraphics[width=\linewidth]{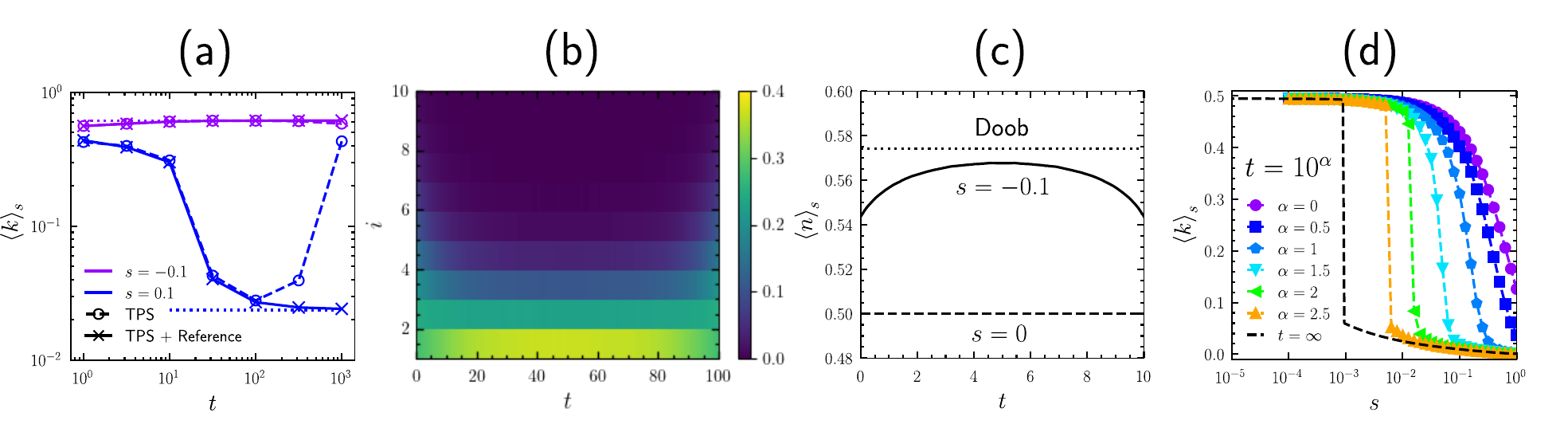}
    \caption{\textbf{Sampling finite time trajectories.} 
    All results are done for the FA model with $c=0.5$.
    (a) The measured dynamical activity $\braket{k}$ as a function of time $t$ for $s=-0.1$ (top) and $s=0.1$ (bottom) for $N_{sp}=10^{6}$ trajectories and system size $N = 40$. The circles show the values obtained via TPS with the normal dynamics, and crosses TPS with the reference dynamics. The dotted lines show the expected value at infinite times.
    (b) The local occupations $\braket{n_{i}}$ as a function of time in the inactive regime $s = 0.1$ and $N = 40$.
    (c) The average excitation density $\braket{n}$ as a function of time in the active regime, $s = -0.1$. The value approaches the expected value in the Doob dynamics (dotted line) in the bulk, but moves towards the equilibrium value ($s=0$, dashed line) at the edges and $N = 40$.
    (d) The dynamical activity as a function of $s$ and time $t$. The data for $t=\infty$ is obtained directly from the MPS, whereas finite $t$ is obtained using TPS. Note that the sharp drop in activity shifts with time. The dynamics are run at the system size $N = 100$.
    }
    \label{fig: finite_time}
\end{figure*}

\subsubsection{Reference dynamics with truncated bond dimensions}

While in the extreme active/inactive limits we can achieve a good MPS description with just a bond dimension of $O(10)$, one may need a bond dimension of $O(100)$ for the more difficult regions such as around $s = 0$ \cite{Banuls2019, Causer2020}.
One reason for the necessity of this high bond dimension could be that the state has longer-ranged spatial correlations. 
Another could be that when one runs the vMPS, we run it against some constraint in the state space. For the FA model, this is the weak constraint that restricts to the connected component of all configuration but the one with $n_{i} = 0$ for all $i$. For the SSEP, we have the stronger constraint that we are within the state space with fixed $N_{p}$ particles.

The goal is to look for a state with a smaller bond dimension than we currently have which still contains all the necessary interactions, but if necessary, discards the information which enforces the constraint. 
Then, by starting our CTMC algorithm in a state which satisfies the constraint, we will automatically enforce it for the rest of the trajectory, as the dynamics keeps the system in the constrained subspace. 

Approximating a TN by another one with a small bond dimension is known as truncation.
For MPS as we use, this can be achieved via a singular value decomposition across each bond, where
only the largest $D'<D$ singular values are kept. We show this truncation in Fig.~\ref{fig: truncated}(a) for SSEP (as this typically requires the largest bond dimension), where we run the vMPS to at least (but higher if required) $D = 50$ to find the state $\ket{\psi_{D}}$, and then truncate to $\ket{\psi_{D'}}$ with the bond dimension $D' < D$.
We measure the truncation error $\varepsilon=1-|\braket{\psi_{D} | \psi_{D'}}|^2$ between the two states, where we assume both are normalised.
We find that when far from the critical point, we can describe the original state to a high accuracy with bond dimensions as small as $D' \sim 20$.
Conversely, we cannot attain the same level of accuracy for $s \sim s_{c}$, where the state exhibits larger amounts of entanglement.

There are multiple reasons that one may want to find a state with a truncated bond dimension. The first is that our Monte Carlo algorithm scales quadratically with the bond dimension - this could hinder the convergence of time-dependent observables at large times, which can require a large sample size to be determined with sufficient accuracy.
For such situations, reducing the scaling of the algorithm would be desired.
Another reasoning could be that we want to investigate a system which requires a higher complexity of TN, such as 2D system with projected entangled pair states (PEPS) \cite{Verstraete2008, Phien2015}. Not only would the scaling of our CTMC algorithm increase, but so would the scaling of the variational algorithm used to find the reference dynamics. In this case, one may not be able to reach a bond dimension large enough to give a desirable variance.

We show the measured dynamical activity for SSEP and the East model (symbols), with a reference dynamics constructed from states with a truncated bond dimension in Fig.~\ref{fig: truncated}(b, c), and compare to the expected result from the non-truncated MPS (dashed line).
Surprinsingly, we find that for the most part, even for bond dimensions as small as $D=2$, we can accurately reproduce the correct dynamical activity for each of the models.
As expected, the truncation struggles mostly around the transition point. Nevertheless, we can achieve good results for the FA (not shown) and East with a truncated bond dimension of $D=4$, and $D=10$ for SSEP.

The calculations done thus far have been with a reference dynamics constructed using a truncated bond dimension without any trajectory re-weighting. In principle, \er{doob_umbrella} is exact and thus allows for further improvements by using the umbrella
\beq
    g(\omega) = e^{-t\theta(s) + \int dt \Delta \hat{R}(\omega)} \text{.}
    \label{reweighting}
\eeq
We implement this re-weighting via transition path sampling (TPS) with the {\em shifting} method, see Refs.~\cite{Bolhuis2002, Oakes2018} for further details.
Figure \ref{fig: truncated}(d) shows the results of this umbrella sampling for the FA model with a $s$ value close to the critical point, $s_{c}$. It is here the discrepancy is the largest, and we can do a more detailed analysis by looking at a larger range of bond dimensions. We see a significant improvement when using the re-weighting factor \er{reweighting}. It might be that we could see further improvements with more TPS iterations.

The main point to take from this is that we are able to achieve accurate results for the dynamical activity (the observable we are tilting) and some local observables with a relatively small bond dimension. This of course comes at a cost however, as when we truncate we discard some of the information that accounts for the long-ranged spatial correlations.
For the case of SSEP, even though apparently we are discarding a large amount of information when truncating (cf.~Fig.~\ref{fig: truncated}(a)), it seems that we keep the relevant information needed to reproduce the correct dynamics, but at the cost of not maintaining the conservation law. We note however that it is possible to explicitly implement the conservation laws in the MPS \cite{Hubig2018ipeps}, but it is not clear how this will affect the quality of the reference dynamics in the CTMC algorithm.

\subsection{Sampling rare events of finite times}

For the previous results, we disregard finite-time effects by considering our sampled trajectories to be a ``slice'' of a larger infinite-time trajectory. We now look to incorporate these effects back into our sampling by considering the full re-weighting factor
\beq
    g(\omega) = \frac{e^{-t\theta(s) + \int dt \Delta \hat{R}(\omega)}}{l_{s}^{\text{ref}}(x_{0}) \, l_{s}^{\text{ref}}(x_{K})} \text{.}
    \label{reweighting_full}
\eeq
Note that previously, for a large bond dimension, the part of \er{reweighting_full} which accounts for the difference in escape rate had a negligible effect, and could be ignored.
This is not always the case here, as the umbrella sampling at the time-edges of the trajectory causes the system to visit configurations which are atypical in the Doob dynamics, and not well described by our MPS approximation.

As a proof of principle, we start by comparing results from TPS with the original dynamics against TPS with the reference dynamics for a small system size $N=40$, and a variety of times, as is shown for the FA model at $s = \pm 0.1$ in Fig.~\ref{fig: finite_time}(a).
For small times, both show excellent agreement.
For large times however, the normal dynamics struggles to correctly account for the expected activity shown by the dotted lines, a result of the exponential time-dependence in \er{expectation_s} (as $\hat{K}$ is time extensive). 
While sampling with our reference dynamics reduces the exponential cost in time, the time-edges still suffer from an exponential sampling difficulty in the system size.
This is most noticeable for the inactive phase, where each model exhibits an exponential localization \cite{Banuls2019, Pancotti2020, Causer2020} at the spatial edge(s) of the system. This causes the $l_{s}^{\text{ref}}(x)$ values to exponentially vary.
Nevertheless, it is still a significant improvement on the previously exponential cost in space, time and $s$.

The average occupations $\braket{n_{i}}_{s}$ (at site $i$) for $s=0.1$ and $t=100$ is shown in Fig.~\ref{fig: finite_time}(b), while Fig.~\ref{fig: finite_time}(c) shows the average excitation density $\braket{n}_{s} = N^{-1}\sum_{i}\braket{n_{i}}_{s}$ for $s=-0.1$ and $t=10$.
It is here the time-edge effects become obvious; we start at a state which lies somewhere between the expected $s=0$ (dashed line) dynamics and the expected long-time dynamics, which depends on the whole spectrum of $\mathbb{W}_{s}$, as well as the total trajectory time.
The system quickly evolves and resembles the Doob dynamics.
Note that at the end of trajectory, it is again described by the original probability vector, as is expected due to the time-symmetry in \er{doob_umbrella}.

Finally, Fig.~\ref{fig: finite_time}(d) shows the average dynamical activity as a function of $s$ and time, $t$.
We show the expected activity in the infinite time limit $t=\infty$ as a black dashed line, and the measured activity for finite times as symbols. Notice that as time decreases, the drop in activity becomes less sharp. Furthermore, the transition from the active to inactive phase happens at decreasing $s$.
While the methods presented here could allow for a detailed investigation into the temporal scaling of the critical point \cite{Appert-Rolland2008, Bodineau2012, Bodineau2012b, Nemoto2017}, doing so for desirable system sizes would be at a large computational cost. We hope to investigate this more extensively using time-evolution methods (see e.g.\cite{Verstraete2008,Schollwoeck2011, Paeckel2019}).

\section{Conclusions}

We have expanded on previous applications of TNs to classical constrained models \cite{Banuls2019, Causer2020, Helms2019, Helms2020}, using the MPS approximation of the leading eigenstates of a tilted stochastic generator in 1D to construct a reference dynamics which well approximates the exact Doob dynamics.
This allows us to (nearly) optimally sample the rare events of 1D constrained systems with just a polynomial cost in both space and time - rather than the exponential cost of most sampling methods. We have demonstrated here the efficiency of this approach by generating tilted trajectory ensembles for the FA and East KCMs and the symmetric simple exclusion process. Our simulations are for sizes and times unprecedented for such large deviation studies. 

Furthermore, our results show that it is possible to obtain an accurate dynamics away from the dynamical transitions of the models we studied with a truncated bond dimension, which enables close to optimal sampling simulations at little cost. 
Further extensions of our work includes generalising our methods to higher dimensions, for example by using two-dimensional variational tensor network techniques, such as PEPS \cite{Verstraete2008, Phien2015} to approximate the leading eigenvectors of 2D classical generators, as is done in Ref.~\cite{Helms2020}. From the associated leading eigenvectors, as we have shown here, we can in turn construct a reference dynamics which is nearly optimal for sampling rare trajectories. 
While PEPS algorithms do not currently allow for bond dimensions comparable to vMPS, they remain a fruitful area of research which is constantly being improved on \cite{Corboz2016, Corboz2016a,Vanderstraeten2016a,Corboz2018prx,Rader2018prx,Vanderstraeten2019,Czarnik2019time, Haghshenas2019, ORouke2020}.
Recent works \cite{casert2020} have shown the effectiveness of using recurrent neural networks (RNN) to approximate the leading eigenstates of tilted generators in two dimensions. The methods presented here can be generalized to RNN to allow for the efficient sampling of 2D rare events.

Another area that deserves exploration is to apply similar TN methods to systems which do not obey detailed balance, and for which their generators cannot be brought to a Hermitian form. While this would damper the effectiveness of variational algorithms, approaches based on time evolution may offer a promising solution (see e.g.\cite{Verstraete2008,Schollwoeck2011,Paeckel2019}).
Such approaches could also offer further insights into intermediate time rare events, where both usual sampling methods and large deviation approaches fall short.
We hope to report on such studies in the near future.

\acknowledgements

We acknowledge financial support from EPSRC Grant no.\ EP/R04421X/1 and the Leverhulme Trust Grant No. RPG-2018-181. 
M.C.B.\ acknowledges support from Deutsche Forschungsgemeinschaft (DFG, German Research Foundation) under Germany's Excellence Strategy -- EXC-2111 -- 390814868.
J.P.G. is grateful to All Souls College, Oxford, for support through a Visiting Fellowship. 
We acknowledge access to the University of Nottingham Augusta HPC service. 
Much of the numerical data  during the later stages of this work was acquired using the ITensor library \cite{itensor}.

\bibliographystyle{apsrev4-1}
%\bibliography{biblo}

\clearpage
\section*{Appendix}
\subsection{Steady state solution in the reference dynamics}
The generator of the reference dynamics defined by \ers{ref_flips}{ref_escapes} can be written as
\beq
    \W^{\text{ref}}_{s} = \sum_{x, x'\neq x} e^{-s} w_{x\to x'} \frac{l_{s}^{\text{ref}}(x')}{l_{s}^{\text{ref}}(x)} \Big[\ket{x'}\bra{x} - \ket{x}\bra{x}\Big] .
    \label{ref_generator}
\eeq
By definition, the stationary state $\ket{\rm ss}^{\text{ref}}_{s} = \sum_{z} P^{\text{ref}}_{s}(z)\ket{z}$ is annihilated by \er{ref_generator}. It follows that
\begin{align}
    \W^{\text{ref}}_{s} \ket{\rm ss}^{\text{ref}}_{s} &= \sum_{x, x'\neq x} e^{-s} w_{x\to x'} \frac{l_{s}^{\text{ref}}(x')}{l_{s}^{\text{ref}}(x)} P^{\text{ref}}_{s}(x) \Big[\ket{x'} - \ket{x}\Big]
    \nonumber
    \\
    & = \sum_{x, x'\neq x}  e^{-s} \Big[w_{x\to x'} \frac{l_{s}^{\text{ref}}(x')}{l_{s}^{\text{ref}}(x)} P^{\text{ref}}_{s}(x)
    \nonumber
    \\
    & \qquad \qquad \qquad - w_{x'\to x} \frac{l_{s}^{\text{ref}}(x)}{l_{s}^{\text{ref}}(x')} P^{\text{ref}}_{s}(x') \Big] \ket{x'}
    \nonumber
    \\
    & = 0 ,
    \label{Appendix_cond}
\end{align}
where we have used a change of variables in the second and third line. Let us assume our original dynamics obeys detailed balance, and that the state space is connected (that is, dynamics is irreducible). Then it follows that if $w_{x\to x'} = 0$, so does $w_{x'\to x}$, in which \er{Appendix_cond} is satisfied. Otherwise, we must have that
\beq
    \frac{P^{\text{ref}}_{s}(x)}{P^{\text{ref}}_{s}(x')} = \frac{w_{x' \to x}}{w_{x \to x'}} \frac{l_{s}^{\text{ref}}(x)^{2}} {l_{s}^{\text{ref}}(x')^{2}} . 
    \label{Appendix_cond2}
\eeq
Given detailed balance we can use a similarity transformation to write the generator in a Hermitian form, cf.~\er{similarity_transformation}. In particular, let us define the diagonal transformation matrix as 
\beq
    Q = \sum_{z}Q(z)\ket{z}\bra{z} .
\eeq
One can easily show that for $\mathbb{H}$ to be Hermitian, we must have
\beq
    \frac{Q(x)^{2}}{Q(x')^{2}} = \frac{w_{x' \to x}}{w_{x \to x'}} .
    \label{Appendix_sub1}
\eeq
Substituting this back into \er{Appendix_cond2}, we find
\beq
    \frac{P^{\text{ref}}_{s}(x)}{P^{\text{ref}}_{s}(x')} = \frac{Q(x)^{2} \, {l_{s}^{\text{ref}}(x)}^{2}} {Q(x')^{2} \, {{l_{s}^{\text{ref}}(x')}^{2}}} . 
\eeq
and it follows the stationary state is given by
\beq
    \ket{\rm ss}^{\text{ref}}_{s} = \sum_{x}{l_{s}^{\text{ref}}(x)}^{2}Q(x)^{2}\ket{x}.
    \label{Appendix_ss}
\eeq
For the case of our MPS dynamics, we defined $\bra{l_{s}} = \bra{\psi_{s}}Q^{-1}$ of our solution $\bra{\psi_{s}}$. It follows that \er{Appendix_ss} can be written as $ \ket{\rm ss}^{\text{ref}}_{s} = \sum_{x}\psi_{s}^{\text{ref}}(x)^{2}\ket{x}$, where $\psi_{s}^{\text{ref}}(x) = \braket{x | \psi_{s}}$.

\subsection{Sampling variance in the reference dynamics}

We start by assuming that we are always at the stationary state of the dynamics. This allows us to calculate the trajectory ensemble average of some observable (per unit time) as the average over all configurations with respect to the stationary state,
\beq
    \braket{\hat{O}}_\text{ref} \equiv \frac{1}{t} \left\langle \int_{0}^{t}dt' \hat{O}(t')\right\rangle_{\text{ref}} = \braket{l_{s}^{\text{ref}} | \hat{O} | r_{s}^{\text{ref}}}.
    \label{ex_traj_config}
\eeq
The aim is to calculate the expectation value and the variance of the time-integrated difference in escape rates (cf. \er{sampling_error_ER}).
Using \er{ex_traj_config}, we can write
\begin{align}
    \braket{\Delta\hat{R}}_{\text{ref}} &= \sum_{x, y\neq x} l_{x} \left( \frac{l_{y}}{l_{x}} e^{-s}\omega_{x\to y} - \omega_{x\to y} \right) r_{x}
    \nonumber
    \\
    &= \sum_{x, y\neq x} l_{y}e^{-s}\omega_{x\to y}r_{x} - l_{x}\omega_{x\to y}r_{x}
    \nonumber
    \\
    &= \braket{l_{s}^{\text{ref}} | \W_{s} | r_{s}^{\text{ref}}} = \theta^{\text{ref}}(s) ,
        \label{avR}
\end{align}
where we have written $l_{x} \equiv l_{s}^{\text{ref}}(x)$ and $r_{x} \equiv r_{s}^{\text{ref}}(x)$ for brevity.
Performing the same calculation for $\Delta \hat{R}^{2}$, we find
\beq
    \braket{\Delta \hat{R}^{2}}_{\text{ref}} = \braket{l_{s}^{\text{ref}} | {\W_{s}}^{2} | r_{s}^{\text{ref}}},
\eeq
giving the variance
\beq
    \Var_{\text{ref}}\Delta\hat{R} \equiv \braket{\Delta \hat{R}^{2}}_{\text{ref}} -  \braket{\Delta\hat{R}}_{\text{ref}}^{2} = \delta E^{2},
    \label{varR}
\eeq
where $\delta E^{2}$ is the measured variance of the MPS used to construct the reference dynamics with respect to the tilted generator (or tilted Hamiltonian).

We are now in a position to estimate the sampling error \er{sampling_error_ER}. From \era{avR}{varR} we have that the integrated difference in escape rate,
\beq
    \int^{t}_{0} dt' \Delta \hat{R}(t') ,
\eeq
has the average 
\beq
    t\braket{\Delta \hat{R}} = t \theta^{\text{ref}}(s) ,
\eeq
and variance
\beq
    t \delta E^{2} .
\eeq
If we also assume this integrated difference to be normally distributed, then we get
\er{sampling_error_ER},
\beq
    \epsilon_{\text{ref}}^{2} = \frac{1}{N_{\text{sp}}} \left[ \frac{\braket{e^{2\int dt' \Delta\hat{R}}}_{\text{ref}}} {\braket{e^{\int dt' \Delta\hat{R}}}_{\text{ref}}^{2}} - 1 \right]
    \approx \frac{e^{t \delta{E}^{2}} - 1}{N_{\text{sp}}} .
\eeq

\end{document}